\documentclass[superscriptaddress,twocolumn,showpacs,preprintnumbers,prl]{revtex4}
\usepackage{times,xspace}
\usepackage{amsbsy,amssymb,amsmath,bm}
\usepackage{graphicx,color,epsfig}
\usepackage{fancyhdr}

\def\bbbc{{\mathchoice {\setbox0=\hbox{$\displaystyle\rm C$}\hbox{\hbox
to0pt{\kern0.4\wd0\vrule height0.9\ht0\hss}\box0}}
{\setbox0=\hbox{$\textstyle\rm C$}\hbox{\hbox
to0pt{\kern0.4\wd0\vrule height0.9\ht0\hss}\box0}}
{\setbox0=\hbox{$\scriptstyle\rm C$}\hbox{\hbox
to0pt{\kern0.4\wd0\vrule height0.9\ht0\hss}\box0}}
{\setbox0=\hbox{$\scriptscriptstyle\rm C$}\hbox{\hbox
to0pt{\kern0.4\wd0\vrule height0.9\ht0\hss}\box0}}}}

\newcommand{\ignore}[1]{}
\newcommand{\mComment}[1]{}
\newcommand{\gComment}[1]{}
\newcommand{\jComment}[1]{}
\newcommand{\rComment}[1]{}
\newcommand{\lComment}[1]{}
\renewcommand{\gComment}[1]{\textcolor{magenta}{Gerardo: #1}}
\newcommand{\LANL}{Theoretical Division, Los Alamos National Laboratory,
Los Alamos, New Mexico 87545}
\newcommand{\IU}{Department of Physics, Indiana University, Bloomington,
IN 47405, USA}

\begin{document}
\title{Field Induced Orbital Antiferromagnetism in Mott Insulators}
\author{K. A. Al-Hassanieh } \affiliation {\LANL}
\author{C. D. Batista} \affiliation {\LANL}
\author{G. Ortiz} \affiliation {\IU}
\author{L. N. Bulaevskii} \affiliation {\LANL}
\date{\today}

\begin{abstract}
We report on a new electromagnetic phenomenon that emerges in Mott 
insulators, i.e., materials that do not conduct electricity because of
strong electronic Coulomb repulsion. The phenomenon manifests as
antiferromagnetic ordering due to orbital electric currents which are
spontaneously generated from the coupling between spin currents and an
external homogenous magnetic field. This novel spin-charge current
effect provides the  mechanism to detect the so far elusive spin
currents by means of unpolarized  neutron scattering, nuclear magnetic
resonance or muon spectroscopy. We illustrate this mechanism by solving
a half-filled Hubbard model on a frustrated ladder, a simple but
nontrivial case of strongly interacting electrons.
\end{abstract}

\pacs{72.80.Sk, 74.25.Ha, 73.22.Gk}
\maketitle

Insulators are generically characterized by the presence of a gap to
charge-carrying  excitations, and classified  according to the origin of
this gap. The charge gap of Mott insulators is driven by intra-atomic
electron-electron Coulomb interactions, and the half-filled Hubbard
Hamiltonian is the minimal model that describes its properties. Mott
insulators exhibit a very broad spectrum  of physical properties due to
the correlated  nature of their electronic localization. While 
electrons are completely localized in the Wannier orbitals of band
insulators, Mott insulators always exhibit a partial electronic
delocalization due to the finitude of the Coulomb repulsion, $U$,
relative to the kinetic energy  or hopping amplitude $t$. 

The combination of partial delocalization with the fermionic nature of
electrons leads to the well known antiferromagnetic exchange, $J$,
between localized spins ${\bf S}_i$. It was shown recently that this  partial
delocalization also leads to electronic charge redistribution 
for a subclass of  bond-ordered spin states \cite{Lev08}.  Even more surprisingly
Mott insulators can also contain a distribution of orbital electric currents
that emerge  for chiral spin orderings \cite{Lev08}.  
These orbital currents are directly
associated with the notion of scalar spin chirality 
\cite{Wen,Kawamura}
\begin{equation}
{\boldsymbol \chi}_{jkl} =  {\bf S}_j \times {\bf S}_k \cdot {\bf S}_l.
\end{equation}

\begin{figure}[thb]
\vspace*{-0.8cm}
\hspace*{-1.0cm}
\includegraphics[angle=0,width=10cm]{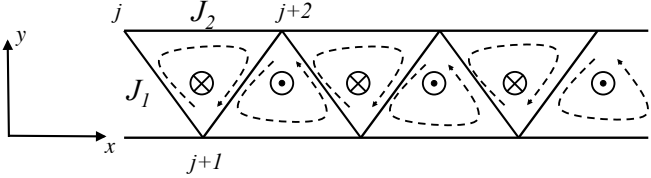}
\vspace{-5.0cm}
\caption{Zigzag ladder. The arrows indicate the circulation of the spin
and the electrical orbital currents that emerge in the ground state for
$J_2 \gg J_1$ and $|B| \lesssim |B_{\rm sat}|$, with $B_{\rm sat}$ the 
saturation field. The  small circles indicate the orientation of the
(staggered) magnetic moments generated by the electric  currents.}
\label{zz}
\end{figure}

The main purpose of this Letter is to demonstrate that a uniform
magnetic field can induce {\it orbital} antiferromagnetism,
i.e., antiferromagnetic ordering due to orbital {\it electric}
currents, by the stabilization of staggered spin currents and the
subsequent interplay between these currents and the applied field. For
this purpose we will consider a Hubbard Hamiltonian on a zig-zag chain
(see Fig.\ref{zz}) with nearest-neighbor  (next-nearest-neighbor)
hopping amplitudes $t_1$ ($t_2$).
The application of a magnetic field $B$ 
increases the uniform magnetization along the field
direction $z$. A spontaneous breaking of the remaining U(1) symmetry of
global spin rotation along the $z$-axis is prevented by quantum
fluctuations (Mermin-Wagner theorem \cite{Mermin66}). This implies the
absence of usual magnetic ordering in the $x$-$y$ plane: $\langle
S^\eta_j \rangle = 0$ with $\eta=x,y$. We will show however that the so
called ``chiral spin orderings'',  both vector chirality ($\langle {\bf
S}_j \times {\bf S}_{j+1} \rangle \neq 0$)  and  scalar spin chirality
($\langle {\boldsymbol \chi}_{ijk} \rangle \neq 0$),  can be induced by
the field.
The field-induced vector chiral
ordering is a remnant of the classical helical order $\langle S^\eta_j
\rangle \neq 0$ that is obtained in the classical (large spin, $S \to
\infty$) limit \cite{Kolezhuk05}. On the other hand, the scalar spin
chirality results from the spontaneous vector spin chirality and the
field induced magnetization, $\langle {\boldsymbol \chi}_{jkl} \rangle
\simeq   \langle{\bf S}_j \times {\bf S}_k \rangle \cdot \langle{\bf
S}_l\rangle$. As we will show below, this connection between
scalar and vector spin chiralities  implies a relation between
electric and spin currents, i.e., a spin-charge current
effect.

Chiral phases (nonzero scalar or vector spin chirality) 
in quantum spin chains were predicted a long time ago and
have been studied for many years. As noted by Villain \cite{Villain78},
the chiral ordering must survive at finite temperature ($T$) in weakly coupled chains,
without the usual helical long-range ordering, since the
chiral correlation length is much longer that the spin correlation
length. In other words, there must be a window of temperatures where
chiral order takes place in absence of helical order.  There are
experimental indications of the existence of such phase in the quasi-one
dimensional organic magnet  Gd(hfac)$_3$NITiPr
\cite{Kolezhuk05,Affronte99}. However, the absence of external physical
fields that couple to non-uniform chiral orderings poses a challenge for
measuring these phases when there is no helical ordering
\cite{Maleyev95}.  {\it By showing that the field-induced vector chirality is
accompanied by a staggered ordering of orbital magnetic moments (orbital
electric currents), we provide a direct way of detecting this exotic
phase by means of unpolarized neutron scattering,  nuclear magnetic
resonance (NMR) or muon spectroscopy ($\mu$-SR).}


We start by considering a  half-filled Hubbard model on the zigzag
ladder with $L$ sites ($0\leq j \leq L-1$) depicted in Fig.\ref{zz},
and  in the presence of magnetic field ${\bf B}= B {\hat {\bf z}}$
\begin{eqnarray}
H = \sum_{j,\nu, \sigma} (t_{\nu} c^{\dagger}_{j\sigma}
c^{\;}_{j+{\nu}\sigma}  +{\rm h.c.}) + \sum_j (U n_{j +} n_{j -} - B 
S^z_j) ,
\end{eqnarray}
where $\nu=1,2$, $\sigma=\pm1$,
$c^{\dagger}_{j\sigma}$($c^{\;}_{j\sigma}$) creates(annihilates) an
electron of spin $\sigma$ at site $j$,
$n_{j\sigma}=c^{\dagger}_{j{\sigma}} c^{\;}_{j{\sigma}}$, $n_j =
\sum_{\sigma} n_{j\sigma}$, and $S^{\eta}_j = \sum_{\alpha \beta}
c^{\dagger}_{j \alpha} \sigma^{\eta}_{\alpha  \beta} c^{\;}_{j\beta} /2$
are the spin-1/2 operators with $\sigma^{\eta}$ the Pauli matrices. We
assume  periodic boundary conditions (PBC), i.e., $L \equiv 0$. In the
large $U/t_{\nu}$ limit, the low-energy spectrum of $H$ is described by  an
effective Heisenberg spin-1/2 Hamiltonian, 
\begin{equation} {\tilde H} = \sum_{j,\nu} J_{\nu} ({\bf S}_{j} \cdot
{\bf S}_{j+\nu}-\frac{1}{4})  - B \sum_j S^z_j ,
\label{heis}
\end{equation}
because the electrons are localized near the lattice sites.  ${\tilde
H}$ is obtained by projecting the original $H$ into the low-energy
subspace ${\cal S}$. This is true in general for any physical quantity,
${\cal A}$, whose effective low-energy operator, ${\tilde {\cal A}}$, is
a function of the spin operators ${\bf S}_{j}$. The expression for
${\tilde {\cal A}}$ is obtained by a canonical transformation that
follows from standard degenerate perturbation theory. The  exchange
constants are $J_{\nu} = 4 t^2_{\nu}/U > 0$.

We are interested in two physical quantities that will be particularly
relevant for this work. These are  the charge and the spin
($z$-component)  current densities in a given bond $\langle jl \rangle$
(with $l=j+\nu$ and $t_{jl}=t_\nu$)
\begin{eqnarray}
{\bf I}^c_{jl} &=&  i  \sum_{\sigma}  (c^{\dagger}_{j\sigma}
c^{\;}_{l\sigma}-c^{\dagger}_{l\sigma} c^{\;}_{j\sigma} )
\frac{et_{jl}}{\hbar}{\hat {\bf r}}_{jl} ,
\nonumber  \\
{\bf I}^s_{jl} &=&  i  \sum_{\sigma}  (c^{\dagger}_{j\sigma}
c^{\;}_{l\sigma}-  c^{\dagger}_{l\sigma} c^{\;}_{j\sigma} )
\frac{\sigma t_{jl}}{\hbar}{\hat {\bf r}}_{jl} ,
\label{current}
\end{eqnarray}
where ${\hat {\bf r}}_{jl}=({\bf r}_l-{\bf r}_j)/|{\bf r}_l-{\bf r}_j|$,
and  ${\bf r}_j$ is the vector position of site $j$. Both are Noether
currents for the charge and spin conservation laws associated with the
corresponding U(1) and SU(2) global symmetries of $H$.
The charge current has a non-zero low-energy effective operator,
${\tilde {\bf I}}^c_{jl}$, whenever the site $j$ belongs to a loop that
is closed by an odd number of hopping terms \cite{Lev08}. Since the
shortest possible loop is a triangle, ${\tilde {\bf I}}^c_{jl}$  is $
{\cal O}(t^{n+1}/U^n)$ with $n$ odd and $n \geq 3$. Therefore, the
lowest order contribution to the effective charge current density
operator is  \cite{Lev08}
\begin{equation}
{\tilde {\bf I}}^c_{jl} = \frac{24 e}{\hbar }\hat {\bf r}_{jl} \sum_{l
\neq j, k}\frac{t_{jl}t_{lk}t_{kj}}{U^{2}} {\boldsymbol \chi}_{jkl} .
\label{chargecurrent}
\end{equation}
Equation (\ref{chargecurrent}) establishes a direct correspondence
between the scalar chiral spin ordering, $\langle {\boldsymbol
\chi}_{ijk}\rangle \neq 0$, and the presence of electric orbital
currents. In other words, most of  the scalar spin orderings will be
accompanied by the emergence of orbital currents.  The effective spin
current density operator is
\begin{eqnarray}
{\tilde {\bf I}}^s_{jl} = \frac{8t^2_{jl}}{U} {\kappa^z_{jl}} \frac{\hat {\bf
r}_{jl}}{\hbar}, 
\;\; {\rm with}\;\;
{\kappa^z_{jl}} =  {\bf S}_j \times {\bf S}_l \cdot {\hat {\bf z}}.
\label{spincurrent}
\end{eqnarray}
This simple expression shows that the spin current is directly
associated to the vector spin chirality. Since  ${\bf I}^s_{jl}$ is even
under a particle-hole transformation ($t_{jl} \to - t_{jl}$, ${\bf S}_j
\to - {\bf S}_j$), the prefactor on the right hand side of Eq.
(\ref{spincurrent}) can only contain even powers of the hopping
amplitudes.

The effective current density operators introduced above are necessary
to characterize the ground state correlations when the magnetic field
approaches its saturation value $B_{\rm sat}$, and $t_2 \gtrsim t_1$.
As we will show below, this ground state exhibits long range order of
spin and charge currents that are roughly proportional to each other
when  $J_2 \gg J_1$, and $|B| \lesssim |B_{\rm sat}|$. 
To understand the  origin of this instability, it is
convenient to write ${\tilde H}$ in terms of new fermionic degrees of
freedom $f_j$ by means of a Jordan-Wigner transformation 
\begin{equation}
S^{+}_j = f^{\dagger}_j  K_j, \;\;\;
S^{-}_j = K_j f^{\;}_j, \;\;\;
{\bar n}_j = S^z_j + \frac{1}{2},
\end{equation}
where ${\bar n}_j =f^{\dagger}_j f^{\;}_j $ and  $K_j =  \prod_{k < j}
(1-2{\bar n}_k)$ is the nonlocal operator that realizes the change
in exchange statistics.  

The new expression of ${\tilde H}$ can be written as
\begin{eqnarray}
{\tilde H} &=& {\tilde H}_{1} +{\tilde H}_{2}- B \sum_l {\bar n}_l , \\
{\tilde H}_{1} \!\!&=& \!\! \frac{J_1}{2} \sum_{l}  (f^{\dagger}_l f^{\;}_{l+1}
+ f^{\dagger}_{l+1} f^{\;}_{l} ) + J_1 \sum_l ({\bar n}_l {\bar n}_{l+1}
-{\bar n}_l) \nonumber \\
{\tilde H}_{2}\!\! &=& \!\! \frac{-J_2}{2} \sum_{l} ({\kappa}^z_l
{\kappa}^z_{l+1}+{\kappa}^z_{l+1} {\kappa}^z_l )+ J_2 \sum_l ({\bar n}_l
{\bar n}_{l+2}- {\bar n}_l) \nonumber,
\end{eqnarray}
where ${\kappa}^z_l \equiv {\kappa}^z_{l l+1}= i (f^{\dagger}_l
f^{\;}_{l+1} - f^{\dagger}_{l+1} f^{\;}_{l})$. Interestingly, we remark
that the first term of ${\tilde H}_2$ is an explicit {\it ferromagnetic}
interaction between spin currents on  adjacent bonds. 
However, a state with net nearest-neighbors spin currents, $\langle
{\kappa}^z_{l} \rangle \neq 0$, can only appear when $J_1 \neq 0$, i.e.,
for a finite coupling between upper and lower chains (see
Fig.\ref{zz}).


To develop some intuition about the role played by $J_1$, it is
convenient to rewrite ${\tilde H}$ in momentum space
\begin{eqnarray}
{\tilde H} = \sum_{k} (\epsilon_k - \mu) a^{\dagger}_k a^{\;}_k +
\frac{1}{2L} \sum_{kqp} v_{pq} \, a^{\dagger}_{q+p} a^{\dagger}_{k-p}
a^{\;}_{k}  a^{\;}_q ,
\label{hefk}
\end{eqnarray}
with $a^{\dagger}_k = \frac{1}{\sqrt{L}} \sum_{j=0}^{L-1} e^{ikj} 
f^{\dagger}_j$, $\epsilon_k = J_1 \cos{k} + J_2 \cos{2k}$, $\mu = B +
J_1 + J_2 $, and $v_{pq}= 2 \epsilon_p - 8 J_2 \cos{(p+2q)}$. The first
term is the non-interacting part of ${\tilde H}$, while  the second
contains the  density-density interactions of ${\tilde H}_{1}$ and
${\tilde H}_{2}$ that lead to the first contribution, $2 \epsilon_p$, 
plus a correlated second-nearest-neighbor hopping that is
contained in the first term of ${\tilde H}_{1}$, and leads to the second
contribution in $v_{pq}$.

For $J_2 > J_1/4$, the fermion dispersion $\epsilon_k$ has two
degenerate minima at $k= \pm Q$, with $\cos{Q} = -J_1/4J_2$, that
correspond to opposite values of ${\kappa}^z_l$.  The saturation field
$B_{\rm sat}=J_1+J_2- \epsilon_{Q}$ ($-B_{\rm sat}$) corresponds to the
critical value of the chemical potential at which the fermion density
$\rho= \langle \bar{n}_j\rangle$ becomes equal to one (zero). From now
on we will assume that the spin system is close to  full polarization.
Since the physics is independent of the sign of $B$ ($B \to -B$ under a
time reversal transformation), we will choose the negative sign,  $B
\gtrsim -B_{\rm sat}$, for the rest of the analysis. For the equivalent
fermionic problem, this condition is equivalent to the dilute density
limit, $\rho \ll 1$ or $\mu \gtrsim \epsilon_{Q}$.

For $J_1=0$, the symmetric and the anti-symmetric sectors generated by
the fermionic operators  $a^{\dagger}_k \pm a^{\dagger}_{k+\pi}$ are
perfectly decoupled. In this case the spin currents are quenched since ${\kappa}^z_l$ can
only connect states on different sectors (it is odd under the
transformation $ k \to k+\pi$).
The situation changes dramatically for $J_1/J_2 \ll 1 $. At low
energies, we can only have fermions $a^{\dagger}_k$  near the two
degenerate minima:  $k \simeq \pm Q$. Since $Q$ is close to $\pi/2$, the
fermions $a^{\dagger}_{\pm Q}$ generate spin currents of nearly maximal
amplitude and opposite signs:  $\sum_l  {\kappa}^z_l
a^{\dagger}_{k}|0\rangle = \sin{k} \; a^{\dagger}_{k}|0\rangle$.  While
the non-interacting part of ${\tilde H}$ favors a ground  state in which
both minima are equally populated (Pauli exclusion principle),
the second term of $v_{pq}$ leads to a nearest-neighbor attraction
between fermions with the same chirality (in the same minimum) and  a
nearest-neighbor repulsion between fermions with opposite vector
chiralities (different minima). This is also  expected from the first
term of ${\tilde H}_{1}$ and it implies the possibility of a chiral
instability  similar to the one originally proposed by Nersesyan and
coauthors \cite{Nersesyan98} for $XY$ zig-zag spin chains  (see also
\cite{Kaburagi99}). In fact, the chiral phase was recently found for $J_2
\gg J_1$ and $|B| \simeq B_{\rm sat}$ \cite{Kolezhuk05} by using the 
mean-field decoupling of the bosonized version of ${\tilde H}$ that was 
originally introduced in Ref. \cite{Nersesyan98} for the anisotropic
case.  On the other hand, a straightforward mean-field approximation to Eq.
(\ref{hefk}) overestimates the stability of the chiral phase (it gives a
wrong density dependence for the energy of the disordered state). Since
mean-field approximations to interacting quasi-one-dimensional systems are
always subjected to scrutiny, it is decisive to have a numerical
confirmation of this phase. By using density-matrix renormalization group (DMRG), McCulloch and
coauthors \cite{McCulloch08} recently found the chiral phase for $J_2=J_1$,
while Okunishi \cite{Okunishi08}  reported a phase diagram that confirms
the existence of a chiral phase for $J_2 \gg J_1$ and $|B| \simeq B_{\rm
sat}$.

We will now derive an important physical consequence of a chiral spin phase
that was overlooked before. To simplify notation we introduce the
definitions: ${\boldsymbol\chi}_{l} \equiv {\boldsymbol \chi}_{l l+1
l+2}$ and ${\bf I}^{\gamma}_{ll+1} \equiv {\bf I}^{\gamma}_{l}$ with
$\gamma= c,s$. The combination of a field induced magnetization, $m_z=\langle
S^z_l \rangle$,  and a net vector spin chirality $\langle \kappa^z_l
\rangle \neq 0$ leads to a non-zero mean value of the {\it scalar spin
chirality}
\begin{equation}
 {\boldsymbol\chi}_{l}  \simeq m_z [ \kappa^z_l  +  \kappa^z_{l+1}  -
 \kappa^z_{l l+2} ].
\label{propor}
\end{equation}
This simple expression shows that, to a good a approximation, scalar and
vector chiralities are proportional to each other, and the uniform
magnetization $m_z$ is the proportionality constant. This relationship
attains a clear physical meaning when we look at Eqs.
(\ref{chargecurrent}) and (\ref{spincurrent}): the field induced  vector
chiral order contains orbital {\it electric currents} that are
proportional to the spin currents
\begin{equation}
{\tilde {\bf I}}^c_{jl} \simeq  \frac{6 e}{U} m_z \sum_{l \neq j, k}
[\frac{t_{jl}t_{lk}}{t_{jk}} {\tilde {\bf I}}^s_{jk}  + 
\frac{t_{jl}t_{jk}}{t_{kl}} {\tilde {\bf I}}^s_{kl} -
\frac{t_{kl}t_{jk}}{t_{jl}} {\tilde {\bf I}}^s_{jl}] . 
\label{spincharge}
\end{equation}
Figure \ref{zz} displays the circulation of electrical orbital currents
(arrows) that are expected for the chiral-ordered ground state of the
ziz-zag chain ($J_2 \gg J_1$ and $|B| \lesssim |B_{\rm sat}|$) 
according to Eq. (\ref{spincharge}). Remarkably, the application of a
{\it uniform} magnetic field induces {\it orbital antiferromagnetism}
via the (dominant) Zeeman coupling to the spin moments. The small
circles in Fig. \ref{zz} indicate the orientation of the staggered
orbital moments that these electric currents generate.

To test the validity of Eq. (\ref{spincharge}) we study the ground state
properties of both $H$ and $\tilde H$  by means of the DMRG method
\cite{White92}. We use PBCs \cite{DMRGnote} to eliminate spurious 
oscillations in the correlation functions due to boundary effects. The
incommensurate nature of the ordered ground state makes the numerical
calculation quite challenging \cite{Aligia00}. In contrast to Ref.
\cite{Okunishi08} we do not include any infinitesimal bias field and
compute chiral-chiral correlators (instead of the order parameter
$\langle \kappa_l \rangle$) to establish chiral long-range order. 
In the case of the Hubbard Hamiltonian
$H$, we solve a chain of $L=64$ sites for $t_2/t_1=1.6$, $U/t_1=20, 24$,
and $30$.  $B$ is chosen such that $2 m_z = 0.875$. The charge-current,
$C^c(r) = \langle I^c_{l}I^c_{l+r}\rangle$, and the spin-current, 
$C^s(r) = \langle I^s_{l}I^s_{l+r}\rangle$, correlation functions are
computed directly in the DMRG ground-state wave function.
\begin{figure}[!htb]
\vspace*{-0.0cm}
\hspace*{-0.5cm}
\includegraphics[angle=0,width=8.5cm]{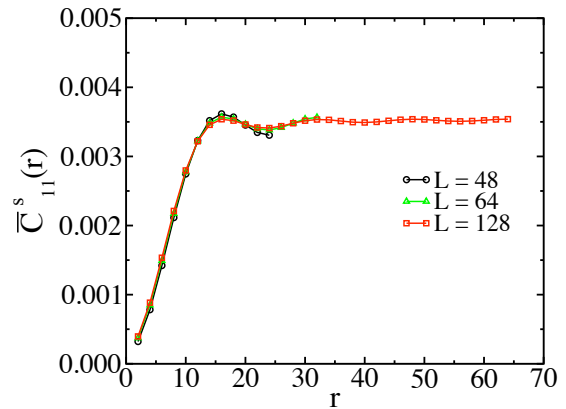}
\vspace{-0.4cm}
\caption{Size dependence of the two-point vector spin chirality
correlator ${\bar C}^s_{nm}(r)$ for $J_2/J_1=2.56$ and $2 m_z = 2
\langle  S^z_l \rangle = 0.875 $.}
\label{correl}
\end{figure}

For the Heisenberg model $\tilde H$, we study chains of lengths
$L=48,64,$ and  $128$ sites. The ratio between the exchange constants is
determined by the ratio between the hopping amplitudes
$J_2/J_1=t_2^2/t_1^2=2.56$ and  $2 m_z = 0.875$.  
In this case we compute the vector spin chirality two-point correlation
functions: ${\bar C}^s_{nm}(r) = \langle \kappa_{ll+n} \kappa_{l+r
l+r+m} \rangle$.  Figure \ref{correl} shows $\bar C^s_{11}(r)$  for
different system sizes.  The finite-size scaling clearly indicates the
presence of long-range vector spin chiral order.

\begin{figure}[!htb]
\vspace*{-0.1cm}
\hspace*{-0.5cm}
\includegraphics[angle=0,width=8.5cm]{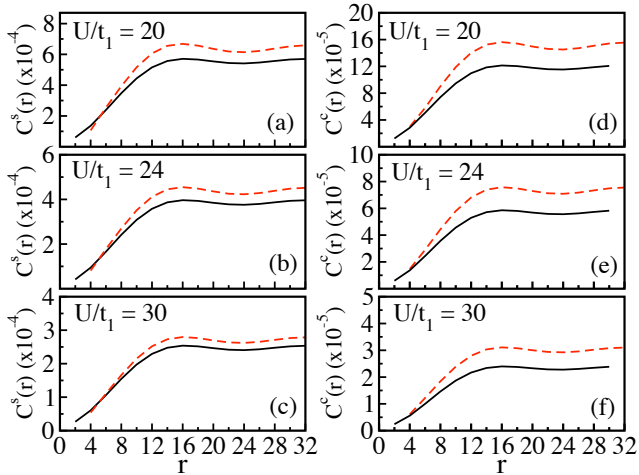}
\vspace{-0.3cm}
\caption{ (a), (b), (c) Charge-current correlation function $C^c(r)$ 
and (d), (e), (f) spin-current correlation function $C^s(r)$ in the
case of  the Hubbard model $H$ (dashed lines) and the Heisenberg model
$\tilde H$ (solid lines).   The results are shown for $L=64$, $t_2/t_1 =
1.6$ ($J_2/J_1 = 2.56$),  $\langle s^z_i\rangle = 0.4375$, and $U/t_1 =
20, 24$, and $30$.}
\label{Hub-Heis}
\end{figure}

According to Eq. (\ref{spincurrent}), the effective spin-current
two-point correlation function is $C^s(r) = [8{t_1^2\over U}]^2 \bar
C^s_{11}(r)$.  At long distances, the effective charge-current
correlation function can be obtained approximately by using Eq.
(\ref{propor})
\begin{eqnarray}
C^c(r) \simeq \tilde{\alpha} \langle {\boldsymbol \chi}_{l} 
{\boldsymbol \chi}_{l+r} \rangle  \simeq 4 \tilde{\alpha} m^2_z [{\bar
C}^s_{11}(r) - {\bar C}^s_{22}(r) + \frac {\bar C^s_{12}(r)}{4}],
\nonumber
\label{approx}
\end{eqnarray}
with $\tilde{\alpha}= 2304 t_1^4t_2^2 / U^4$. Figures \ref{Hub-Heis}(a),
(b) and (c) compares the spin-current two-point correlators computed
with the Heisenberg (${\tilde H}$) and the Hubbard ($H$) models for
three different values of $U/t_1=20, 24$ and $30$. 

As expected the results for the two models show better agreement as $U$ 
increases. According to Eq. (\ref{approx}), the long-range ordered
spin-currents (see  Fig. \ref{correl})  must lead to long-range ordered
orbital electric currents. Figures \ref{Hub-Heis}(a), (b) and (c) show a
comparison between $C^c(r) = \langle I^c_{l}I^c_{l+r}\rangle$ computed
with the Hubbard Hamiltonian ($H$) for the same three different values
of $U$, and the right-hand side of Eq. (\ref{approx}) computed  with the
Heisenberg model ${\tilde H}$. The good agreement between both curves
confirms that the spin and the  electric currents are approximately
proportional to each other, with the proportionality constant linear in
$m_z$.  In other words, the ordering of spin currents is accompanied by
ordering of orbital electric currents for non-zero $m_z$ as follows from
Eq. (\ref{propor}).

The resulting staggered configuration of orbital magnetic moments
implies that the ground state under  consideration is a field-induced
orbital antiferromagnet. We note that the orbital moments are located at
the centers of the triangles (see Fig. \ref{zz}), while the spin moments
are obviously located at the corners (or sites of the lattice).  The
values of the orbital currents that we obtained for $U/t_1 =20$ (see
Fig. \ref{Hub-Heis}(d)) are of order 0.01 $t_1$. The corresponding
orbital magnetic moment is a few percent of a Bohr magneton for $t_1
\sim 1$eV.  In contrast to the spin currents $\langle \kappa^z_l
\rangle$, these orbital magnetic moments can be measured with different
experimental techniques such as neutron scattering, NMR or $\mu$-SR.
This provides a simple way of detecting this exotic spin ordering in
real materials. Moreover, notice that the magnitude of the magnetic
moment would be much larger if the chiral phase remains stable in the
intermediate coupling  regime $U \gtrsim t_{\nu}$.

To induce an electric current out of an spontaneously  generated spin
current (or vice-versa) one needs an externally applied magnetic field
(or any other field that is odd under a time reversal transformation) 
since both currents have opposite parity under time reversal.  
This symmetry consideration has to be complemented by a microscopic
mechanism that determines the magnitude of the spin-charge current
conversion. The possibility of having orbital electric currents in
frustrated Mott insulators \cite{Lev08} is the key for finding and
quantifying such mechanism. This spin-charge current effect cannot be
associated with a magnetoelectric response because the spin current does
not couple directly to any electric or magnetic fields. On the other
hand, the induced orbital electric currents can lead to small
magnetoelectric effects. For instance, the application of an electric 
potential between the upper and lower chains (see Fig. \ref{zz}) will
lead to a  net orbital magnetization because the magnetic moments in the
lower and upper triangles will not be compensated any longer.

This work was carried out under the  auspices of the National Nuclear
Security Administration of the U.S. Department  of Energy at Los Alamos
National Laboratory under Contract No. DE-AC52-06NA25396.


\begin{thebibliography}{12}


\bibitem{Lev08}
L. N. Bulaevskii, C. D. Batista, M. Mostovoy, and D. Khomskii, 
Phys. Rev. B {\bf 78}, 024402 (2008).

\bibitem{Wen}
X. G. Wen, F. Wilczek, and A. Zee, Phys. Rev. B {\bf 39}, 11413 (1989).

\bibitem{Kawamura} 
H. Kawamura, Phys. Rev. Lett. {\bf 68}, 3785 (1992).

\bibitem{Mermin66}
N. D. Mermin and H. Wagner, Phys. Rev. Lett. {\bf 17}, 1133 (1966).

\bibitem{Kolezhuk05}
A. Kolezhuk and T. Vekua, Phys. Rev. B {\bf 72}, 094424 (2005).

\bibitem{Villain78}
J. Villain, Ann. Isr. Phys. Soc. {\bf 2}, 565 (1978).

\bibitem{Affronte99}
M. Affronte {\it et al.}, Phys. Rev. B {\bf 59}, 6282 (1999).

\bibitem{Maleyev95}
S. V. Maleyev {\it et al.}, J. Phys. Condens. Matter {\bf 10}, 951
(1998); S. V. Maleyev, Phys. Rev. Lett. {\bf 75}, 4682 (1995). 

 \bibitem{Nersesyan98}
A. A. Nersesyan, A. O. Gogolin, and F. H. L. Essler, Phys. Rev. Lett.
{\bf 81}, 910 (1998). 

\bibitem{Kaburagi99}
M. Kaburagi, H. Kawamura, and T. Hikihara, J. Phys. Soc. Jpn. {\bf 68},
3185 (1999); T. Hikihara {\it et al.}, J. Phys. Soc. Jpn. {\bf 69}
259 (2000); Y. Nishiyama, Eur. Phys. J. B {\bf 17}, 295 (2000). 

\bibitem{McCulloch08} 
I. P. McCulloch {\it et al.}, Phys. Rev. B {\bf 77}, 094404 (2008).

\bibitem{Okunishi08}
K. Okunishi, J. Phys. Soc. Jpn. {\bf 77}, 114004 (2008).

\bibitem{White92} 
S. R. White, Phys. Rev. Lett. {\bf 69}, 2863 (1992); 
Phys. Rev. B {\bf 48}, 10345 (1993); K. Hallberg, Adv. Phys. {\bf 55}, 
477 (2006); U. Schollw\"ock, Rev. Mod. Phys. {\bf 77}, 259 (2005); 
A.F. Albuquerque {\it et al.}, J. Mag. Mag. Mat. {\bf 310}, 1187 (2006).

\bibitem{DMRGnote} 
In the finite-system step, we keep up to $M=1200$  states and perform up
to 20 sweeps.  The weight of the discarded states  is of order
$10^{-11}$ or smaller.

\bibitem{Aligia00}
A. A. Aligia, C. D. Batista, and F. H. L. Essler, Phys. Rev. B {\bf 62},
3259 (2000).
















\end{thebibliography}
\end{document}